\documentclass[journal]{IEEEtran}
\usepackage{cite}

\usepackage{graphicx}
\usepackage[cmex10]{amsmath}
\usepackage{algorithmic}

\usepackage{array}
\usepackage[caption=false,font=footnotesize]{subfig}
\usepackage{tikz}
\usetikzlibrary{positioning}

\usepackage[shortlabels]{enumitem}

\begin{document}
%
\title{Kinetic Compressive Sensing}
%
%
%

\author{Michele~Scipioni,~\IEEEmembership{Student Member,~IEEE,}
        Maria~F.~Santarelli,
        Luigi~Landini,
        Ciprian~Catana,
        Douglas~N.~Greve,
        Julie~C.~Price, and
        Stefano~Pedemonte

\thanks{M.~Scipioni and L.~Landini are with the Department of Information Engineering, University of Pisa, Pisa, Italy (email: \textit{michele.scipioni@ing.unipi.it})}%
\thanks{M.F.~Santarelli and L.~Landini are with the Institute of Clinical Physiology, CNR, Pisa, Italy}%
\thanks{M.~Scipioni, C.~Catana, D.N.~Greve, J.C.~Price, and S.~Pedemonte are with the Athinoula A. Martinos Center for Biomedical Imaging, Boston, MA, USA.}%
\thanks{C.~Catana, D.N.~Greve, J.C.~Price, and S.~Pedemonte are with Harvard Medical School, Boston, MA, USA.}%
\thanks{S.~Pedemonte is with the Center for Clinical Data Science, MGH, Boston, MA, USA}%

}

\IEEEpubid{978-1-5386-2282-7/17/\$31.00~\copyright~2017 IEEE}


\maketitle

\begin{abstract}
Parametric images provide insight into the spatial distribution of physiological parameters, but they are often extremely noisy, due to low SNR of tomographic data. Direct estimation from projections allows accurate noise modeling, improving the results of post-reconstruction fitting. We propose a method, which we name kinetic compressive sensing (KCS), based on a hierarchical Bayesian model and on a novel reconstruction algorithm, that encodes sparsity of kinetic parameters. Parametric maps are reconstructed by maximizing the joint probability, with an Iterated Conditional Modes (ICM) approach, alternating the optimization of activity time series (OS-MAP-OSL), and kinetic parameters (MAP-LM). We evaluated the proposed algorithm on a simulated dynamic phantom: a bias/variance study confirmed how direct estimates can improve the quality of parametric maps over a post-reconstruction fitting, and showed how the novel sparsity prior can further reduce their variance, without affecting bias. Real FDG PET human brain data (Siemens mMR, 40min) images were also processed. Results enforced how the proposed KCS-regularized direct method can produce spatially coherent images and parametric maps, with lower spatial noise and better tissue contrast. A GPU-based open source implementation of the algorithm is provided.
\end{abstract}

\begin{IEEEkeywords}
parametric images, PET, compartmental models, compressive sensing, hierarchical Bayesian model, sparsity, Markov Random Field, FDG, GPU
\end{IEEEkeywords}

%
\IEEEpeerreviewmaketitle

\section{Introduction}
%
%
%
%
\IEEEPARstart{K}{inetic} analysis of medical imaging data is a standard approach to characterize the body's physiological processes, allowing \textit{in vivo} studies of metabolic patterns, perfusion, blood flow, ligand-protein binding interactions, response to pharmacological challenges, and gene expression. 

There are two approaches for deriving kinetic parameters: region-of-interest (ROI) modeling and parametric imaging. The ROI based approach fits a kinetic model to the average time activity curve (TAC) of a selected ROI: it is easy to implement and has low computational cost. In contrast, parametric imaging estimates kinetic parameters for every voxel, providing richer kinetic data and enabling the visualization of the spatial distribution of physiological parameters. However, parametric images are often extremely noisy, as a reflection of the noisy measurement of tomographic data. 

In conventional indirect kinetic analysis, the parametric maps are estimated by reconstructing a time series of tomographic images, then fitting a compartmental kinetic model. Direct estimation of parametric images \cite{reader2014} from raw projection data, which has become of interest with the availability of more powerful computers and of GPUs, allows accurate noise modeling and has been shown to offer better image quality than conventional indirect methods. Despite the improvement introduced by direct kinetic modeling, parametric images remain still extremely noisy. 

In this work, we use a novel approach for integrating information derived from kinetic modeling into the dynamic reconstruction in the form of a regularization prior, in order to reduce the effect of fitting errors on image reconstruction and convergence. Moreover,
in order to extract from the noisy PET measurements spatially coherent parametric images, we propose a maximum-a-posteriori version of the standard Levemberg-Marquardt nonlinear least square optimization, where we use an Huber prior to enforce an hypothesis of sparsity of the spatial derivatives of the kinetic maps. 

The method, which we name kinetic compressive sensing (KCS), is based on a hierarchical Bayesian model, and it can be described as an Ordered-Subsets Maximum-A-Posteriori One-Step-Late reconstruction algorithm with Kinetics-derived Prior (OS-MAP-OSL-KP) regularization term. In this article, we describe experiments with both simulated data and real [18F]FDG PET data. We also provide a GPU-based open source implementation of the reconstruction algorithm \cite{gpuKMfit}. 

\IEEEpubidadjcol

\section{Methods}
The method is based on a hierarchical Bayesian model, from which we derived an optimization algorithm for the direct estimation of the kinetic parameters, which includes sparsity constraints. We provide an efficient GPU implementation of the projection and back-projection operations, as well as of the parametric maps estimation, based on the analytic evaluation of the gradient of the log joint posterior of the Bayesian model.

\subsection{Hierarchical Bayesian model}
The overall model is composed of three parts: 
\begin{enumerate}[a)]
    \item the model of the acquisition system;
    \item the kinetic model;
    \item a model to promote sparsity of the kinetic maps' derivatives. 
\end{enumerate}

\begin{figure}[!t]
\centering
\includegraphics[width=3.3in]{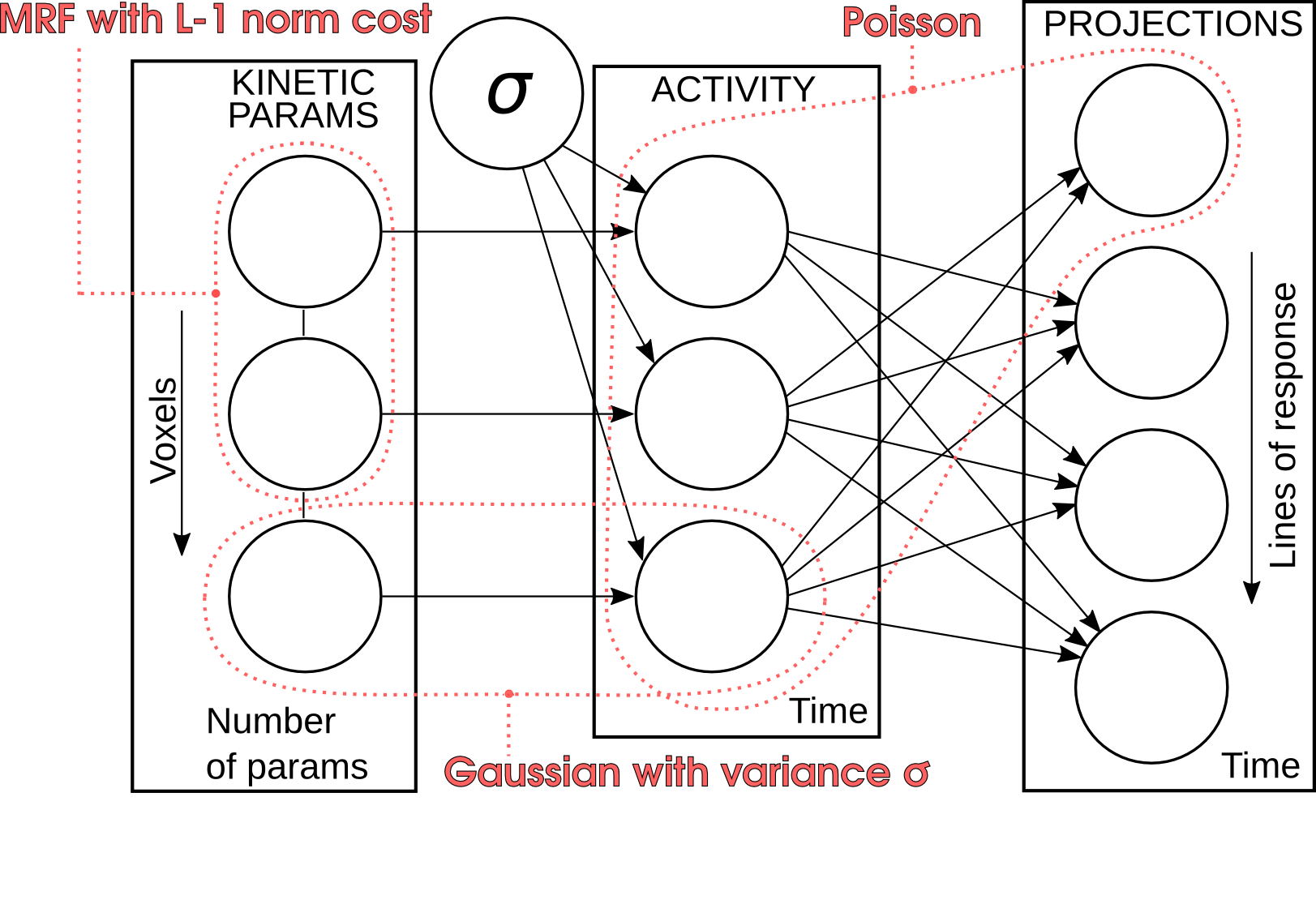}
\vspace{-1.5em}
\caption{Hierarchical Bayesian Model. From left to right: the kinetic model treats each voxel as a realization of a hidden dynamic process; the sparsity-inducing prior acts voxel-wise on the parametric maps as an MRF with L1-norm cost function; the model of the acquisition process links image domain and measured projections.}
\label{bayesian_model}
\end{figure}

Reading the graph in fig.\ref{bayesian_model} from right to left, we first find the level of projection domain. This is linked to the activity image domain by the model of the acquisition system: it consists of the ordinary Poisson model, incorporating all effects of attenuation, scatter and randoms. Since we are working with dynamic 4D datasets, each voxel has a relevant time activity curve (TAC): the kinetic model encodes the assumption that the voxel intensities are noisy realizations of a hidden dynamic process, modeled in this case using a multi-compartmental model. The one-to-one connection between elements of the activity domain and elements of the kinetic parameters domain is the result of model fitting: each TAC is linked to a set of parameters, and so each layer of the kinetic parameters domain can be seen as a parametric map. The sparsity-inducing prior is introduced at this level as a Markov Random Field (MRF) with an L1-norm cost function. 

\subsection{Image reconstruction with kinetics-derived prior}

In this work we chose to incorporate time information derived from kinetic modeling as a regularization term to help reducing fluctuations and noise in successive frames of our time series, which are common if we reconstruct each one of them independently.

In the general MAP framework, we want to maximize the log-posterior distribution of our measured data $s$, with respect to the image estimate $x$, and subject to a regularization term $U(x)$. This usually means looking for $\hat{x}$ that maximizes the log-posterior probability:

\begin{equation}
    l(s|x) = logP(s|x)-\beta U(x) + const
    \label{logpriorGibbs}
\end{equation}

where $P(s|x)$ is the conventional Poisson likelihood, and $\beta$ is a positive scaling factor.

As shown by Green \cite{OSL}, this problem can be solved using the one-step-late (OSL) approach for an iterative update of the MAP estimate:

\begin{equation}
    x_{vm}^{new} =  \frac{x_{vm}^{old}}{\sum_{d} p_{vd} - \beta\frac{\partial U(x)}{\partial x_{vm}} \bigg\rvert_{x_{vm}=x_{vm}^{old}}   } \sum_{d} \frac{p_{vd} s_{dm}}{\sum_{v}p_{vd} x_{vm}^{old} }
\label{reconstruction}
\end{equation}

where $v$ is the voxel index, $m$ is the time frame number, a single bin $d$ of the projection data measured at time $m$ is represented as $s_{dm}$, and $p_{vd}$ represents an element of the system matrix, modeling the contribution of voxel $v$ to projection bin $d$. The OSL strategy involves calculation of the derivative of $U(x)$, in order to update the image vector $x$ from the old estimate.

Regarding the choice of the potential function $U(x)$, our proposal is specifically designed for 4D dynamic acquisitions because the information we want to introduce as a prior is aimed to suppress the temporal fluctuations (i.e. noise) in each voxel's TAC due to the independent reconstruction of each time frame. This issue has been addressed by other direct reconstruction methods by forcing the new image update to start from the output of the kinetic modeling step computed on the previous 4D estimate. On the contrary, we thought of introducing a time constraint through a Gibbs prior as a gentler way to constrain the reconstruction, while still being as close as possible to the measurement.

To do this, we chose our potential function $U(x)$ so to enforce similarity between correspondent time points between the reconstructed time series $x_{v:}^{new}$ and the model simulation $F(\mathbf{\xi}_{v}^{old})_:$ computed at the previous step. As for standard OSL approach, in order to do this we have to approximate $x_{v:}^{new}$ as $x_{v:}^{old}$.

\begin{equation}
    U(x) \propto \frac{[F(\mathbf{\xi}_{v}^{old})_m - x_{vm}^{old}]^2}{2\sigma^2}
    \label{prior}
\end{equation}

where $\mathbf{\xi}_{v}^{old}$ represent the set of kinetic parameters estimated for voxel $v$, and $F(\mathbf{\xi}_{v}^{old})_m$ is the modeled TAC for voxel $v$ at time point $m$. 

To implement the aforementioned OS-MAP-OSL-KP reconstruction algorithm (\ref{reconstruction})-(\ref{prior}), we used the in-house developed \textit{Occiput.io} \cite{occiput} software, which uses GPU parallel computation for the operations of projection and back-projection to speed up the reconstruction process. 

\subsection{Update of parametric maps}
To update the prior in (\ref{prior}) after each iteration of OS-MAP-OSL-KP (\ref{reconstruction}), we need to voxelwise fit the kinetic model to the intermediate image update $x_{vm}^{new}$. This step is usual for most common nested direct reconstruction algorithms. In this work we focused on tackling the two most known issues of the standard approach:
\begin{enumerate}[a)]
    \item the execution time needed to loop through each voxel's TAC to fit the model;
    \item the low SNR of the kinetic maps due to different convergence points for each voxel, when we treat each one of them as an independent optimization problem.
\end{enumerate}

\paragraph{Fast parallel GPU implementation}

To address the issue of long execution time, we chose to move the problem of fitting a non linear compartment model to every voxel in our image from CPU to GPU. GPUs are single-instruction-multiple-data (SIMD) devices. This means they are very efficient at performing the same algorithmic steps, in parallel, on large sets of data. In this work, we exploited this property of GPU computing to develop a set of CUDA kernels to evaluate the time integrals of the kinetic model and its derivatives with respect to the unknown parameters.

We also developed a CUDA implementation of a Maximum-a-Posteriori Levemberg-Marquardt (MAP-LM) algorithm for nonlinear least square optimization, based on \textit{cuBLAS} batched matrix operations. An additional speed-up was obtained by avoiding numeric integration, deriving the analytic expression of the model equations and their derivatives, following an idea proposed in \cite{miccai2015}. 

In short, the update of parametric maps follows the following equation:
\begin{equation}
\begin{split}
    \xi_v^{new} = \xi_v^{old} + \{ (J_v^T J_v + \lambda I)^{-1}&[ J_v^T ( x_v - F(\mathbf{\xi_v^{old}})) + \\ 
    & - \beta\frac{\partial H(\xi_v)}{\partial \xi_v} ] \}
\end{split}
    \label{model fitting}
\end{equation}

where, for the current voxel $v$ assigned to a specific GPU thread, $\xi_v$ is the estimated parameter set, $J_v$ is Jacobian matrix, and $x_v$ is the reconstructed TAC. $H(\xi_v)$ is a potential function used as spatial regularization term for MAP fitting of the model to the current voxel, whose derivation is described in the next paragraph.

\paragraph{Derivation of the regularization term for MAP-LM optimization}
The modeling step in (\ref{model fitting}) is an iterative process of its own. Thanks to the parallel GPU implementation, we are now able to synchronize the update of the fitting for all the voxel so that, after each iteration of (\ref{model fitting}) we have an intermediate estimate of the parametric maps we can use to derive the prior term $H(\xi_v)$.

In an attempt to reduce parametric maps' spatial noise, and to guide the convergence of voxels with similar kinetic to the same minima (i.e. parameter sets), we decided to impose an \textit{a priori} assumption to the LM optimization based on a roughness penalty. We chose a Huber function: 

\begin{equation}
H(\xi) \propto 
     \begin{cases}
        t^2                           & \mbox{if } |t| \leq \delta \\
        \delta |t|-\frac{\delta^2}{2} & \mbox{if } |t| > \delta 
     \end{cases}
\end{equation}

\begin{equation}
\frac{\partial H(\xi)}{\partial \xi} = 
     \begin{cases}
        -\delta    & \mbox{if } t < -\delta \\
        t          & \mbox{if } -\delta \leq t \leq \delta  \\
        -\delta    & \mbox{if } t > \delta 
     \end{cases}
\label{huber_deriv}
\end{equation}

where

\begin{equation}
    t = \xi_j-\xi_k \quad \mbox{    for    } j,k \in \Omega
\end{equation}

The idea is to rely on information derived from differences in voxel values computed on a local neighborhood, with a local penalty that penalizes large differences (i.e. borders) less than a quadratic penalty.

The actual filtering kernel $H(\xi)$ used in (\ref{model fitting}) is derived from the Markov Random Field (MRF) prior conditioned on a non-interacting line-site model, presented in fig.\ref{bayesian_model}, which uses small cliques of $3 \times 3$ neighboring voxels. The convolution between Huber kernel and current estimate of parametric maps, $\xi$, is computed in a highly parallel fashion thanks to the development of a dedicated CUDA kernel running on GPU.

\subsection{Summary of the algorithm}

\begin{figure}[!t]
\centering
\includegraphics[width=2.5in]{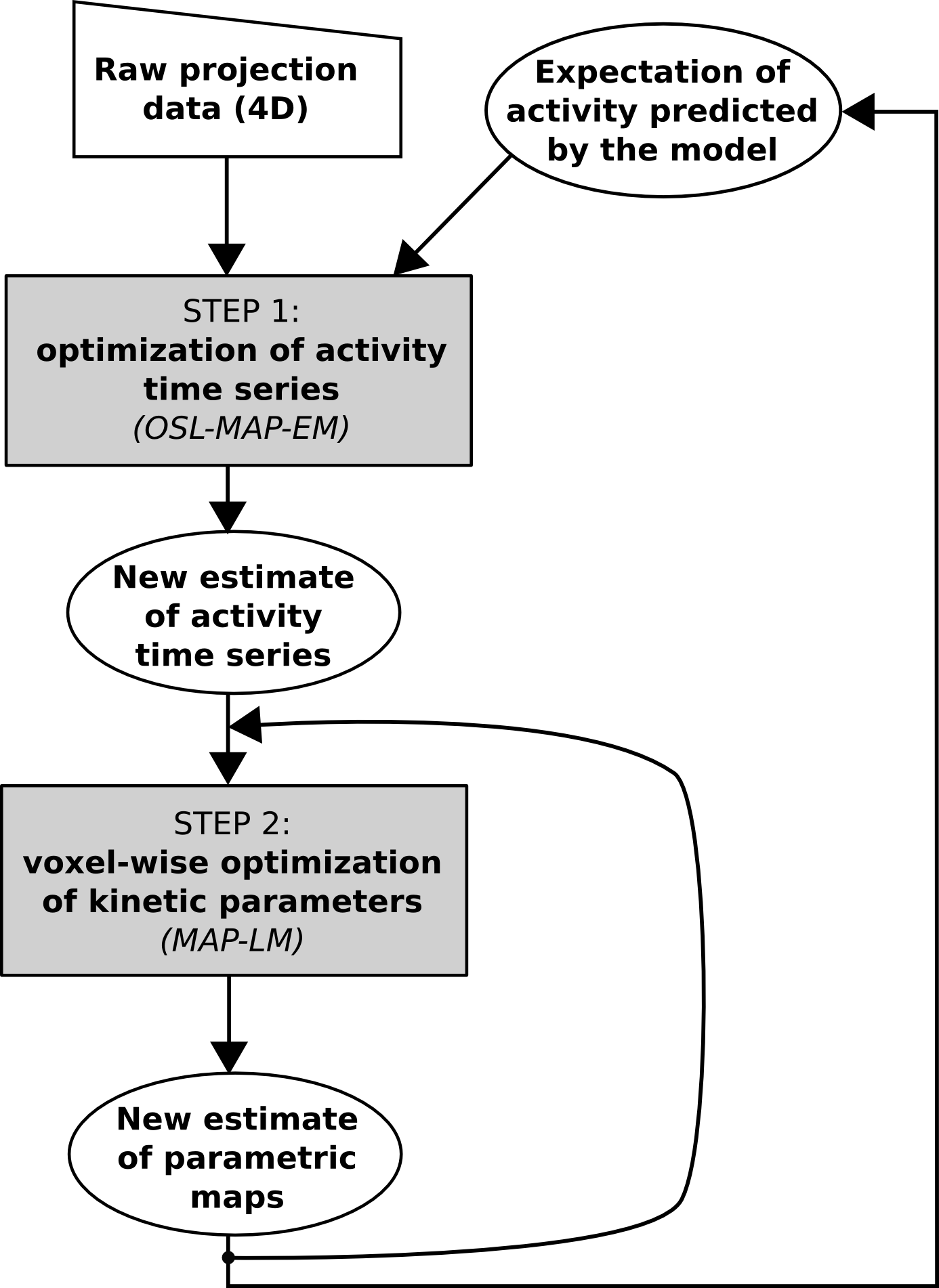}
\caption{Algorithm flow chart: we alternate between the image reconstruction via the OS-MAP-OSL-KP algorithm, and the GPU parallel implementation of parametric mapping via MAP-LM.}
\label{flow_chart}
\vspace{0.0em}
\end{figure}

The parametric maps are reconstructed by maximizing the joint probability of all unknown parameters in the graph in Fig.\ref{bayesian_model} by means of the iterated conditional modes (ICM) algorithm \cite{ICM}. As shown in fig.\ref{flow_chart}, the algorithm consists of alternating the optimization of the activity time series and of the kinetic parameters.

\begin{enumerate}[a)]
    \item given an updated estimate of the kinetic maps, we use it to produce a new estimate of the image time series via OS-MAP-OSL-KP algorithm;
    \item given the new estimated activity time series, we updated the parametric maps via MAP-LM optimization. 
\end{enumerate}


\section{Simulation}

\subsection{Simulation setup}
\begin{figure}[t!]
    \centering
    \includegraphics[width=1.5in]{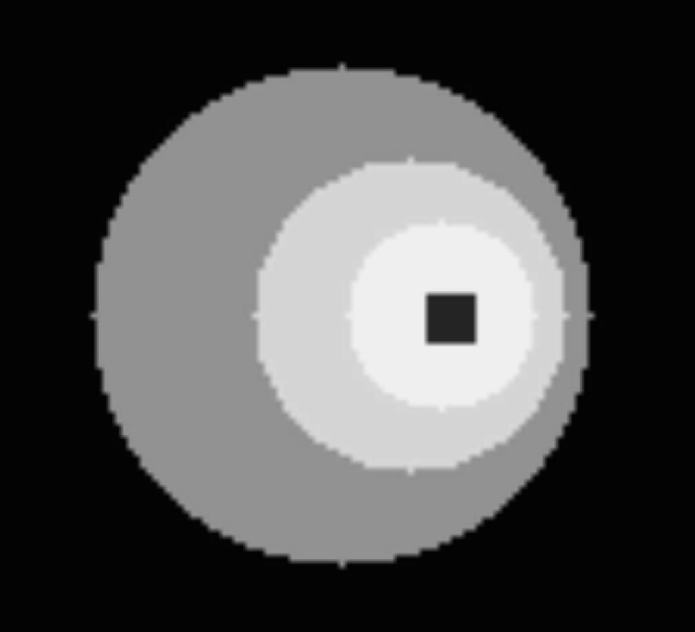}
    \caption{Simulated geometrical phantom}
    \label{fig:phantom}
\end{figure}

To assess the performance of the parallel GPU implementation, and the effect of the sparsity-inducing regularization of the kinetic map estimation, we realized a Monte Carlo (MC) simulation with 100 noisy realizations using the phantom in Fig.\ref{fig:phantom}. 

The kinetic behavior of the three main regions has been simulated using a 2-tissues irreversible compartment model, using the parameters shown in table \ref{table_simul}. The square region in the center is chosen as to emulate a blood input region. 

The non-uniform time scheme used for the simulation covers 40 minutes of acquisition with 24 time frames, and it matches the protocol used for the following human data study: \newline 
\begin{equation}
    [12\times 10s; 2\times 30s; 4\times 60s; 1\times 120s; 4\times 300s; 1\times 600s]
    \label{time_scheme}
\end{equation}

\begin{table}[!t]
\renewcommand{\arraystretch}{1.5}
\caption{Kinetic parameters used in simulation}
\label{table_simul}
\centering
\begin{tabular}{c c c c c }
\hline
Region & $f_v$ (\%) & $K_1$ ($ml/g/min$) & $k_2$ ($min^{-1}$) & $k_3$ ($min^{-1}$) \\
\hline
Blood  & 1.0 & 0.0 & 0.0 & 0.0 \\
Inner  & 0.13 & 0.75 & 0.35 & 0.031 \\
Middle & 0.116 & 0.62 & 0.30 & 0.026 \\
Outer  & 0.0985 & 0.51 & 0.27 & 0.018 \\
\hline
\end{tabular}
\end{table}

\subsection{Simulation results}
We compared the results of three different methods:
\begin{enumerate}[a)]
    \item indirect reconstruction (INDIRECT): OSEM and post-fitting of kinetic model;
    \item direct reconstruction (DIRECT): OS-MAP-OSL-KP an presented in (\ref{reconstruction}), but without using a spatial regularization in (\ref{model fitting});
    \item direct reconstruction with kinetic compressive sensing (KCS DIRECT): like the previous one but with the addition of the prior term in (\ref{model fitting}) defined as in (\ref{huber_deriv}).
\end{enumerate}

\begin{figure}[b!]
    \centering
    \includegraphics[width=2.7in]{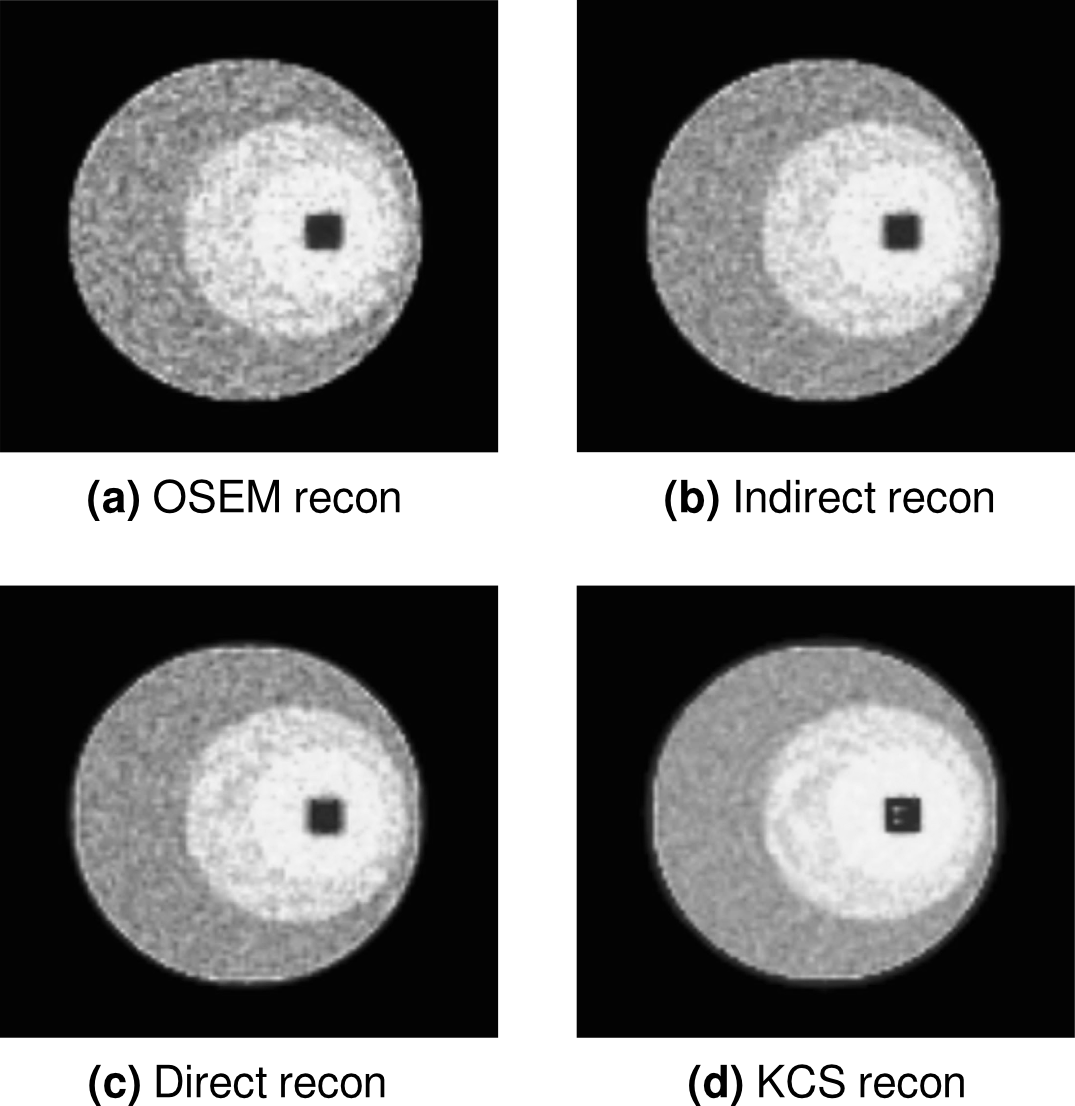}
    \caption{Simulation study: late time frame reconstruction performed using (a) OSEM reconstruction, (b) INDIRECT post-reconstruction fitting, (c) DIRECT reconstruction, and (d) KCS DIRECT reconstruction.}
    \label{sim_fig}
    \vspace{0.0em}
\end{figure}

Fig. \ref{sim_fig} shows the reconstruction of a late time frame. It is easy to recognize a first reduction in voxel-by-voxel variance when the kinetic model is used to regularize the reconstruction (c), which is further reduced when the sparsity assumption of the spatial derivatives of the parametric maps is enforced (d). 

\begin{figure}[!t]
    \centering
    \includegraphics[width=2.7in,height=2.7in]{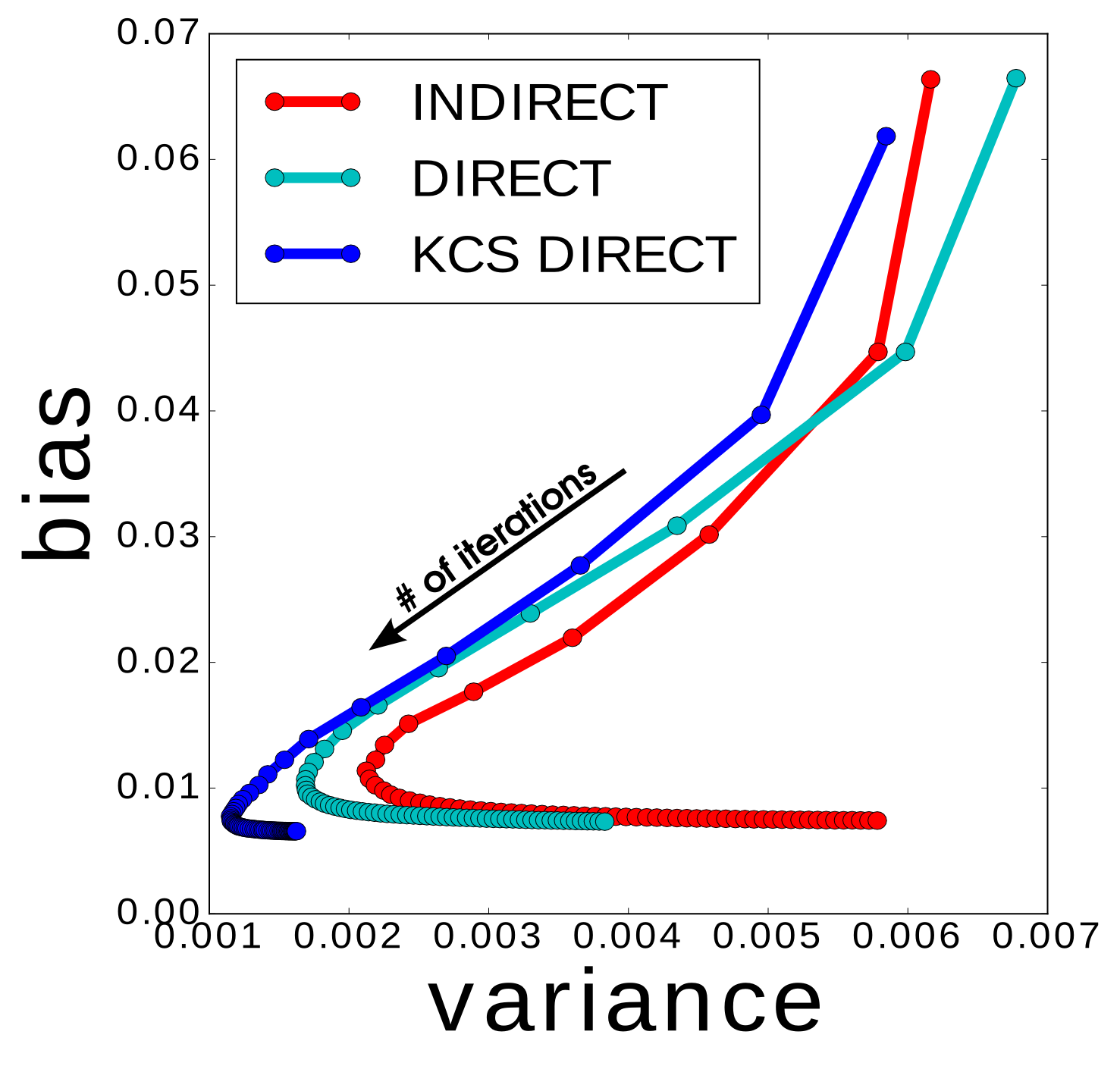}
    \caption{Bias/variance plot of the estimated parametric maps, for the different methods.}
    \label{fig:bias_var_plot}
\end{figure}

The bias/variance plot in fig. \ref{fig:bias_var_plot} shows how a direct approach improves the quality of the estimate of parametric maps, with respect to the results provided by a standard indirect post-reconstruction fitting, but also how the novel sparsity constraint is able to further reduce the variance of the produced parametric maps, without affecting, if not decreasing, the bias, also for a high number of reconstruction iterations.


\section{Human Data}

\subsection{Dataset}
The same three methods tested by simulation (conventional indirect, direct, and KCS) were applied to a [\textsuperscript{18}F]FDG brain PET scan. The data were acquired on a Siemens mMR PET-MR scanner, recording a 40 minute 3D listmode dataset.

The raw data were then rebinned into sinograms, using the same time scheme described in (\ref{time_scheme}) for the simulation.

Given the kinetic behavior of the radiotracer under study in brain, during the reconstruction and modeling phase we used the same 2-tissue irreversible compartment model adopted for the simulation.  

\subsection{Results}

\begin{figure}[t!]
    \centering
    \includegraphics[width=3.45in]{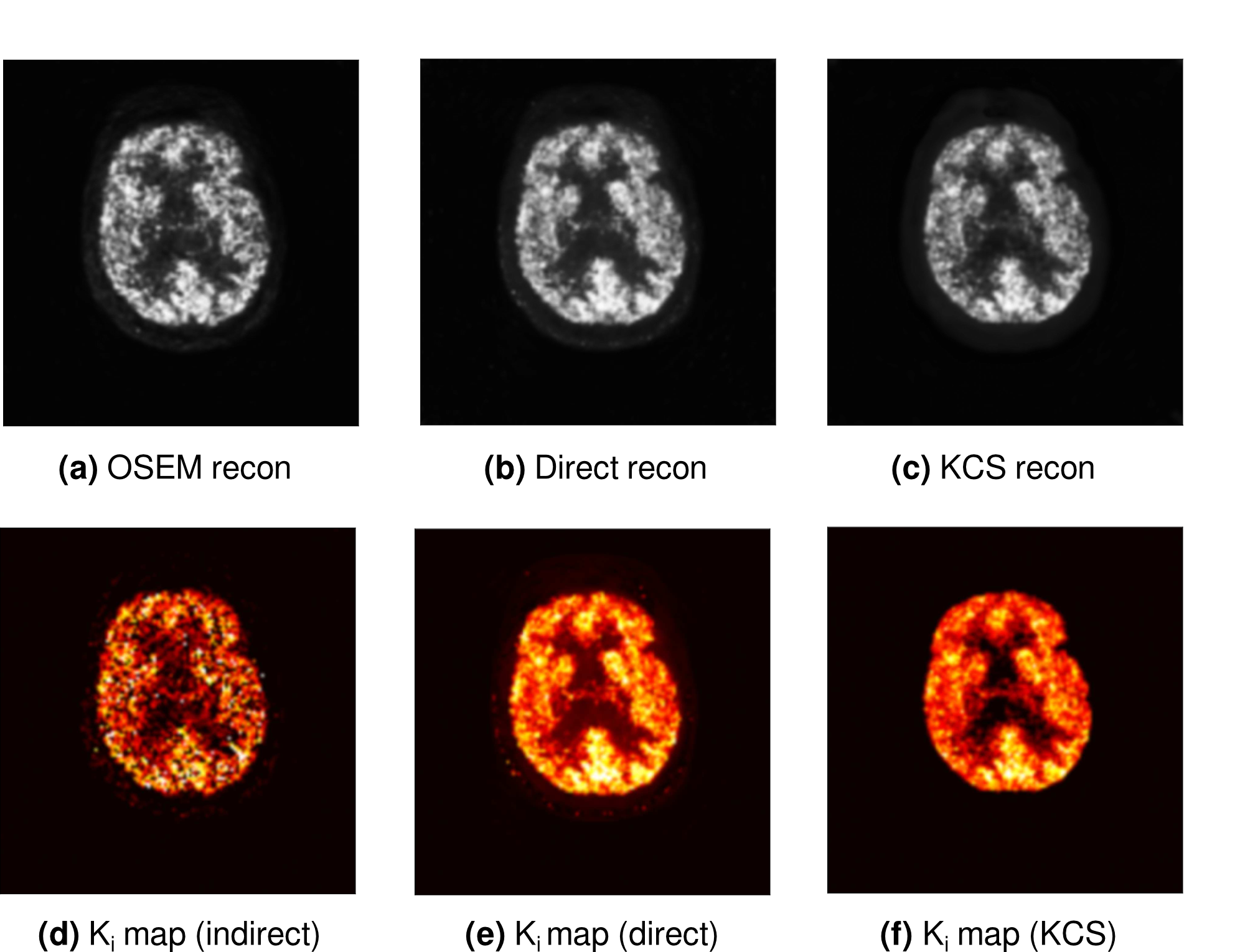}
    \caption{Human data results. Top row: the last reconstructed time frame generated by standard OSEM reconstruction (a), OS-MAP-OSL-KP direct approach, which uses the kinetic model as a guidance during the reconstruction (b), and the KCS-regularized version of direct reconstruction (c). Bottom row: the $K_i$ parametric map produced by each of the three tested methods.}
    \label{real_fig}
\end{figure}

Fig.\ref{real_fig} shows a comparison of the results obtained using the three different methods, in terms of images reconstruction (top row) and $K_i$ (net uptake rate) map estimate (bottom row). Both perspectives convey the same message: the proposed method (c-f) is able to produce spatially coherent images with lower noise, above all with respect to the indirect estimate (a-d), and better tissue contrast, also compared to a direct method without the spatial KCS regularization (b-e).

It is important to take into account how in this work we decided to test the algorithm with a bi-compartmental model to describe tracer uptake in tissue. This means that the final output, in terms of parametric maps, is a set of maps, one for each micro parameter of the model ($K_1, k_2, k_3, f_v$). 

We chose here to show the map of the macro-parameter $K_i$, derived as $K_i = K_1 k_3 / (k_2+k_3)$, also to make it easier to compare the results with other similar methods using linear models (i.e. Patlak).

\vspace{0.5cm}

\section{Conclusion}
The simulation study demonstrated that the proposed method of introducing a sparsity-inducing prior in a direct reconstruction framework can help in producing high-quality images and parametric maps, that are both amenable for display and also quantitatively more accurate than what a post-reconstruction fitting and an unconstrained direct reconstruction can achieve, from a bias/variance point of view. 

This method appears to be promising as a feasible approach to applying kinetic modeling to very large 4D clinical datasets with a reduced computational cost, thanks to the parallel GPU implementation based on the analytic expression of the kinetic model and its derivatives. 

Future studies will focus on evaluating the performance of the proposed OS-MAP-OSL-KP (\ref{reconstruction}) reconstruction algorithm with respect to other known approaches for direct reconstruction \cite{reader2014}, with and without the spatial KCS prior presented in this work.


Moreover, the proposed KCS algorithm could be adapted to other applications, such as kinetic modeling of dynamic contrast enhanced (DCE) MRI and CT.


%

\vspace{0.5cm}
\section*{Acknowledgment}
The Tesla K20 used for this research was donated by the NVIDIA Corporation.

\bibliographystyle{IEEEtran}
\bibliography{biblio}

\end{document}